\documentstyle[preprint,prl,aps,epsf]{revtex}
\begin{document}

\tightenlines

\preprint{\vbox{\hbox{BNL-64627}
\hbox{PRINCETON/HEP/97-11}
\hbox{TRI--PP--97--27}
\hbox{KEK Preprint 97-114}}}

\draft
\title{Observation of the Decay $K^+\!\rightarrow\!\pi^+\mu^+\mu^-$}


\author{S. Adler, M.S. Atiya, I-H. Chiang, J.S. Frank, J.S. Haggerty,
  T.F. Kycia, K.K. Li, \\ L.S. Littenberg, 
  A. Sambamurti\cite{AKS}, A. Stevens, R.C. Strand, and C. Witzig}
\address{
  Brookhaven National Laboratory, Upton, New York 11973
}

\author{W.C. Louis}
\address{
  Medium Energy Physics Division, Los Alamos National Laboratory, \\
Los Alamos, New Mexico 87545
}

\author{D.S. Akerib\cite{DSA}, M. Ardebili\cite{MA}, M. Convery\cite{MC}, 
M.M. Ito\cite{MMI}, 
D.R. Marlow, R. A. McPherson\cite{RMcP}, \\ 
P.D. Meyers, M.A. Selen\cite{MAS}, F.C. Shoemaker, and A.J.S. Smith}
\address{
  Joseph Henry Laboratories, Princeton University,
Princeton, New Jersey 08544
}

\author{E.W. Blackmore, D.A. Bryman, L. Felawka,
  A. Konaka, 
  Y. Kuno\cite{YK}, J.A. Macdonald, \\ T. Numao, 
P. Padley\cite{PP},
  J.-M. Poutissou,
  R. Poutissou, J. Roy\cite{JR}, and A.S. Turcot\cite{AT}}
\address{
  TRIUMF, Vancouver, British Columbia, Canada, V6T 2A3
}

\author{P. Kitching, T. Nakano\cite{TN}, M. Rozon\cite{MR}, and
  R. Soluk}
\address{
 Centre for Subatomic Research, University of Alberta, Edmonton, Alberta, 
Canada, T6G 2N5
}


\date{October 14, 1997}
\maketitle

\begin{abstract}
We have observed the rare decay $K^{+}\!\rightarrow\!\pi^{+}\mu^{+}\mu^{-}$
and measured the branching ratio
$\Gamma(K^{+}\!\rightarrow\!\pi^{+}\mu^{+}\mu^{-})/\Gamma(K^{+}\!\rightarrow\!
\hbox{all}) =
(5.0 \pm 0.4^{\rm stat} \pm 0.7^{\rm syst} \pm 0.6^{\rm th}) \times 10^{-8}$.
We compare this result with predictions from chiral perturbation theory and 
estimates based on the decay $K^+ \!\rightarrow\! \pi^+ e^+ e^-$.
\end{abstract}



We report the observation of the rare decay 
$K^{+}\!\rightarrow\!\pi^{+}\mu^{+}\mu^{-}$
and the measurement of the branching ratio 
$\Gamma(K^{+}\!\rightarrow\!\pi^{+}\mu^{+}\mu^{-})/\Gamma(K^{+}\!\rightarrow\!
\hbox{all})$.
The elec\-tro\-mag\-netically-induced semi-leptonic weak processes 
$K^+ \!\rightarrow\!
\pi^+ l^+ l^-$ ($l=e,\,\mu$) 
have been the subject of considerable theoretical study
for over 40 years~\cite{vains}.  
Though gauge theory
calculations in the free quark approximation\cite{GL} gave a result
in rough agreement with the subsequent observation of 
$K^+ \!\rightarrow\! \pi^+
e^+ e^-$~\cite{bloch}, the inclusion of QCD corrections
\cite{vains,GAW} showed this agreement to be fortuitous
and established that these decays are, in fact, long-distance dominated.
Ecker, Pich, and de Rafael\cite{ecker} first applied the
techniques of chiral perturbation theory (ChPT) to this problem,
and most recent discussion has been conducted in
this language\cite{except}.  In \cite{ecker}, the rate and dilepton
invariant mass spectrum of $K^+ \!\rightarrow\! \pi^+ l^+ l^-$ 
are calculated in
terms of a single unknown parameter, $w_+$.  Consequently, this
parameter also controls the ratio of rates 
$\Gamma(K^+ \!\rightarrow\! \pi^+ \mu^+
\mu^-)/\Gamma(K^+ \!\rightarrow\! \pi^+ e^+ e^-)$.  Calculations of $w_+$ in
various models\cite{ecker,models} have ranged from 0.49 to 2.04.

In the first experiment able to probe this picture, Alliegro
{\it et al.}\cite{alliegro}, made a combined fit to the branching
ratio and spectrum  of $K^+ \!\rightarrow\! \pi^+ e^+ e^-$, 
obtaining $B(K^+ \!\rightarrow\! \pi^+ e^+ e^-) = (2.99
\pm 0.22^{\rm stat} \pm 0.14^{\rm syst}) \times 10^{-7}$ 
and $w_+ = 0.89 {+0.24 \atop{-0.14}}$.  The
value of $w_+$ determined from the branching ratio alone is $1.20 \pm 0.04$,
a little more than $1 \sigma$ above the value obtained from
the simultaneous fit.  
Our previous search for the decay mode 
$K^{+}\!\rightarrow\!\pi^{+}\mu^{+}\mu^{-}$ set a limit of 
$B(K^{+}\!\rightarrow\!\pi^{+}\mu^{+}\mu^{-} ) \le 2.3 \times 10^{-7}$ 
(90\% confidence level) \cite{e787pmm}.
The measurement of $B(K^{+}\!\rightarrow\!\pi^{+}\mu^{+}\mu^{-} )$ allows 
further tests of the ChPT picture.
$K^+ \!\rightarrow\! \pi^+ l^+l^-$ decays are also of 
interest for the light they can shed
on the closely related decays $K_S \!\rightarrow\! \pi^0 l^+ l^-$,
which are important in isolating the
contribution of direct CP-violation in the decay 
$K_L \!\rightarrow\! \pi^0 l^+
l^-$\cite{DG}.

The experiment reported here
was carried out at the Brookhaven Alternating Gradient
Synchrotron from 1989 to 1991 using the E-787 apparatus 
described in Ref. \cite{e787nim}.
An 800 MeV$/c$ $K^{+}$ beam is tagged by a \v Cerenkov counter
and stopped in a scintillating fiber target.
Decay particles from the target are detected
in nested cylindrical layers 
of trigger scintillators, 
a tracking drift chamber, a ``range stack'' of scintillator layers
read out with 500-MHz Transient Digitizers (TD),
and 14 radiation lengths of Pb-scintillator photon detector.
Azimuthally, the trigger counters, range stack, and photon detector are
divided into 6, 24, and 48 sectors, respectively.
Both ends of the cylinder are instrumented with 12-radiation-length 
Pb-scintillator ``endcap'' photon detectors.
The entire apparatus is in a 1-T solenoidal magnetic field.

In each 1.5-s beam spill, approximately $3\times 10^5$ kaons enter
the stopping target.
Requiring a
delay $\, >$1.5 ns between the entering and decay particles
ensures that the kaons stopped in the target before decaying.
The trigger then selects decays in which two or three charged particles 
reach the range stack with $\left|\cos\theta\right|<0.5$
($\theta$ is the polar angle with respect to the beam direction)
and none penetrate beyond 14 cm
at $90^{\circ}$.
Events with more than 5 (10) MeV in the cylindrical (end cap) 
photon detectors
or more than 5 MeV in the outer region of the range stack are eliminated
by the trigger.
These requirements are aimed at the major backgrounds:
$K^+\!\rightarrow\! \pi^+\pi^0$ 
or $K^+\!\rightarrow\!\pi^0\mu^+\nu_\mu$, followed
by $\pi^0\!\rightarrow\!\gamma e^+e^-$ (Dalitz decay) or by 
$\pi^0\!\rightarrow\!\gamma\gamma$ with
$\gamma\!\rightarrow\! e^+e^-$; and
$K^+\!\rightarrow\! \pi^+\pi^+\pi^-$.
In total, we accumulated about $6 \times 10^{6}$ triggers
from $3\times 10^{11}$ stopped kaons.

We search these data for $K^{+}\!\rightarrow\!\pi^{+}\mu^{+}\mu^{-}$ 
events using two
experimental signatures analyzed in parallel.
One signature 
requires complete reconstruction of the three-track 
$K^{+}\!\rightarrow\!\pi^{+}\mu^{+}\mu^{-}$\ events
in the drift chamber and target.
Not only can momentum conservation be ensured for such events, but 
particle identification techniques such as $dE/dx$ measured in the drift
chamber can be applied to all three tracks.
The other signature
uses the fact that the spectrometer, in which accepted charged tracks penetrate
little material other than scintillator, is a hermetic detector of kinetic 
energy.
The segmentation of the kaon stopping target allows separation
of the kaon signal from the decay-product signals, and a 
$K^{+}\!\rightarrow\!\pi^{+}\mu^{+}\mu^{-}$\ event
can be recognized by its total kinetic energy of 143 MeV, even if 
one of the tracks is not reconstructed outside the target.  
This ``two-track'' signature 
\cite{mathesis}
is important because
Monte Carlo simulation indicates that only 10\% of 
$K^{+}\!\rightarrow\!\pi^{+}\mu^{+}\mu^{-}$\ events that pass
the trigger have
a third track that leaves the target.  

The preliminary event selections for
the two signatures are similar.  
First, reconstructed tracks are
required to originate near the kaon track in the target.
Next, no event is accepted having
a track with momentum
greater than the $K^{+}\!\rightarrow\!\pi^{+}\mu^{+}\mu^{-}$\ 
endpoint of 172 MeV$/c$ --- most other kaon
decays have a higher-momentum track.
Events with photons or showering electrons are rejected if they have
energy deposited in the photon detectors or in range stack
modules isolated from the charged tracks (typical threshold $\sim 1$ MeV).
To avoid excessive loss of acceptance from the
high accidental rates in these detectors, only depositions within
1.0--3.5 ns of the charged tracks (depending on analysis and detector) 
are used.
Further rejection of events with electrons comes from demanding that the total
energy measured in the 
inner range-stack layers be less than
150 (120) MeV for the three- (two-)track analysis.

At this point, the major backgrounds are 
$K^{+}\!\rightarrow\! \pi^{+}\pi^{-} e^{+} \nu$,
$K^+\!\rightarrow\!\pi^0\mu^+\nu$ with Dalitz decay, and 
$K^{+}\!\rightarrow\!\pi^{+}\pi^{+}\pi^{-}$,
the last of these entering only the two-track sample.
Dramatic reduction of these backgrounds occurs when we require that
two positively-charged tracks reach the range stack.
For the first two backgrounds, this requires that
the positron reach the range stack, and all the electron identification
techniques of both analyses can be brought to bear on it.
It is difficult for both $\pi^+$'s in
$K^{+}\!\rightarrow\!\pi^{+}\pi^{+}\pi^{-}$ events to reach the range stack.
If they do, the $\pi^-$ is confined to the target, increasing the
effectiveness of the final kinematic requirements.

For each track reaching the range stack, the momentum measured in the drift
chamber can be combined with the kinetic energy measured in the range
stack to give the mass of the particle.
The r.m.s. resolution on this mass is 10--20 MeV$/c^2$ for pions and muons 
in the momentum range of interest.
Both analyses reject electrons by requiring $m>60$ MeV$/c^2$ for any 
tracks reaching the range stack.

\begin{figure}
\leavevmode
\hfil\epsfysize=4.0in\epsfbox{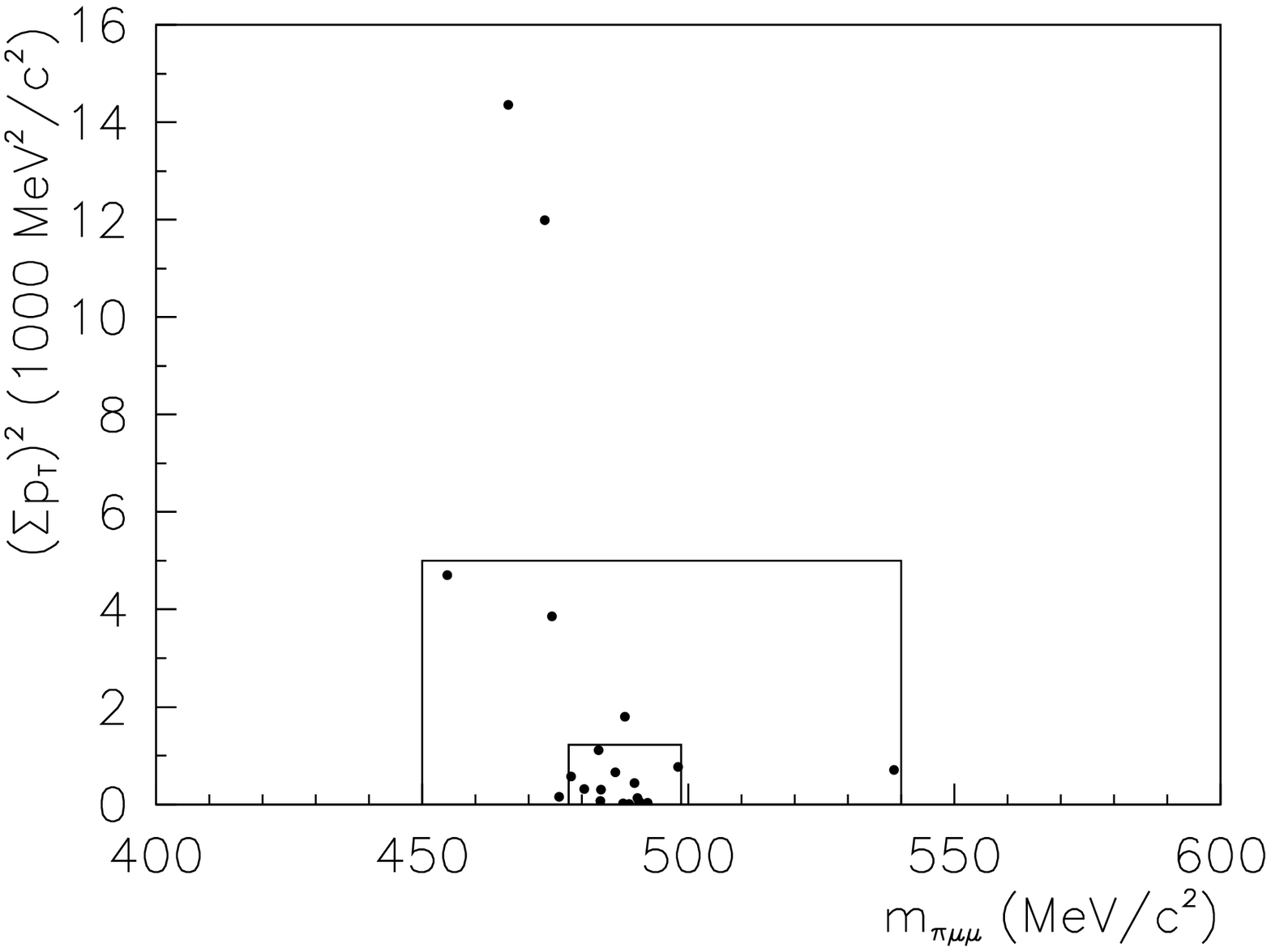}\hfil
\caption{Kinematics of final sample in the three-track analysis.
$\left|\Sigma {\bf p}_{T}\right|$ is the total transverse momentum
(near zero for signal), and $m_{\pi\mu\mu}$ is the three-track
invariant mass ($\approx m_K$ for signal). The inner and outer
boxes are for determining signal and background (see text).
There are two events with 
$\left|\Sigma {\bf p}_{T}\right|^2>16000$ MeV$^2/c^2$.}
\label{final90}
\end{figure}

The final requirements of the two analyses differ sharply.
In the three-track analysis, 
$K^{+}\!\rightarrow\! \pi^{+}\pi^{-} e^{+} \nu$ decays dominate the background.
Some positrons are rejected by restricting the number of range stack counters
allowed on each track, further suppressing showers.
To eliminate remaining background events with electrons or positrons,
information from drift chamber $dE/dx$, $dE/dx$ in the first layer of
the range stack, and time of flight from the trigger counter to the range stack
are combined into a likelihood whose acceptance and rejection are measured
on samples of known $e^{\pm}$ and $\pi^+$ \cite{idsamp}.
Requiring a 90\%-confidence-level elimination of electrons
results in a
sample of 22 events. Fig.~\ref{final90} shows
the total transverse momentum,
$\left|\Sigma {\bf p}_{T}\right|$,
calculated by extrapolating the three tracks back to the decay vertex and 
correcting for energy loss in the target, versus the three-track
invariant mass of this sample.
The assignment of $\pi^{+}$ or $\mu^{+}$ is chosen to minimize 
$\left|\Sigma {\bf p}_{T}\right|$ (the mass enters only in corrections for
propagation in the target), 
but $\pi^+$ and $\mu^+$ are not otherwise distinguished.
Monte Carlo simulation of 
$K^{+}\!\rightarrow\!\pi^{+}\mu^{+}\mu^{-}$ indicates that about 80\% of the
events that remain after preliminary event selection should
appear in the signal box, the smaller of the two boxes in the figure.
There are 13 events in this box.

The background in the signal box is estimated by counting the number of 
events in a
background box surrounding the signal box (see Fig.~\ref{final90})
and 
using a Monte Carlo prediction of the ratio of background events
in the two regions,
taking into account the fraction of signal events outside the signal box.
The background estimated this way is $2.4 \pm 2.2$ events.
This is consistent with an estimate using
trigger counter $dE/dx$, which is not an ingredient of the
particle identification likelihood,
to count events in the signal box with an $e^+$.
It is also consistent with an absolutely-normalized Monte Carlo calculation.
The net signal from the three-track analysis is thus $10.6\pm 4.7$ events.

The final requirements in the two-track analysis impose momentum 
conservation. 
For the small minority of events in the sample with three reconstructed tracks
in the drift chamber, we demand that the vector sum of the momenta be
less than 60 MeV$/c$.
For the rest of the events, the expected energy of the third track
and its angle in the plane 
transverse to the beam are calculated from the two reconstructed tracks,
assuming $K^{+}\!\rightarrow\!\pi^{+}\mu^{+}\mu^{-}.$
Requiring that the ``stub'' of the third track seen in the target 
be within 0.9 radians of the expected direction rejects
$K^{+}\!\rightarrow\! \pi^{+}\pi^{-} e^{+} \nu$, which has a missing neutrino.
Requiring that the energy of the
stub be within 25 MeV of its expected kinetic energy rejects both
$K^{+}\!\rightarrow\! \pi^{+}\pi^{-} e^{+} \nu$ and 
$K^{+}\!\rightarrow\!\pi^{+}\pi^{+}\pi^{-}$ because the $\pi^-$ 
in these events usually deposits
substantial energy after nuclear capture.

\begin{figure}
\leavevmode
\hfil\epsfysize=4.0in\epsfbox{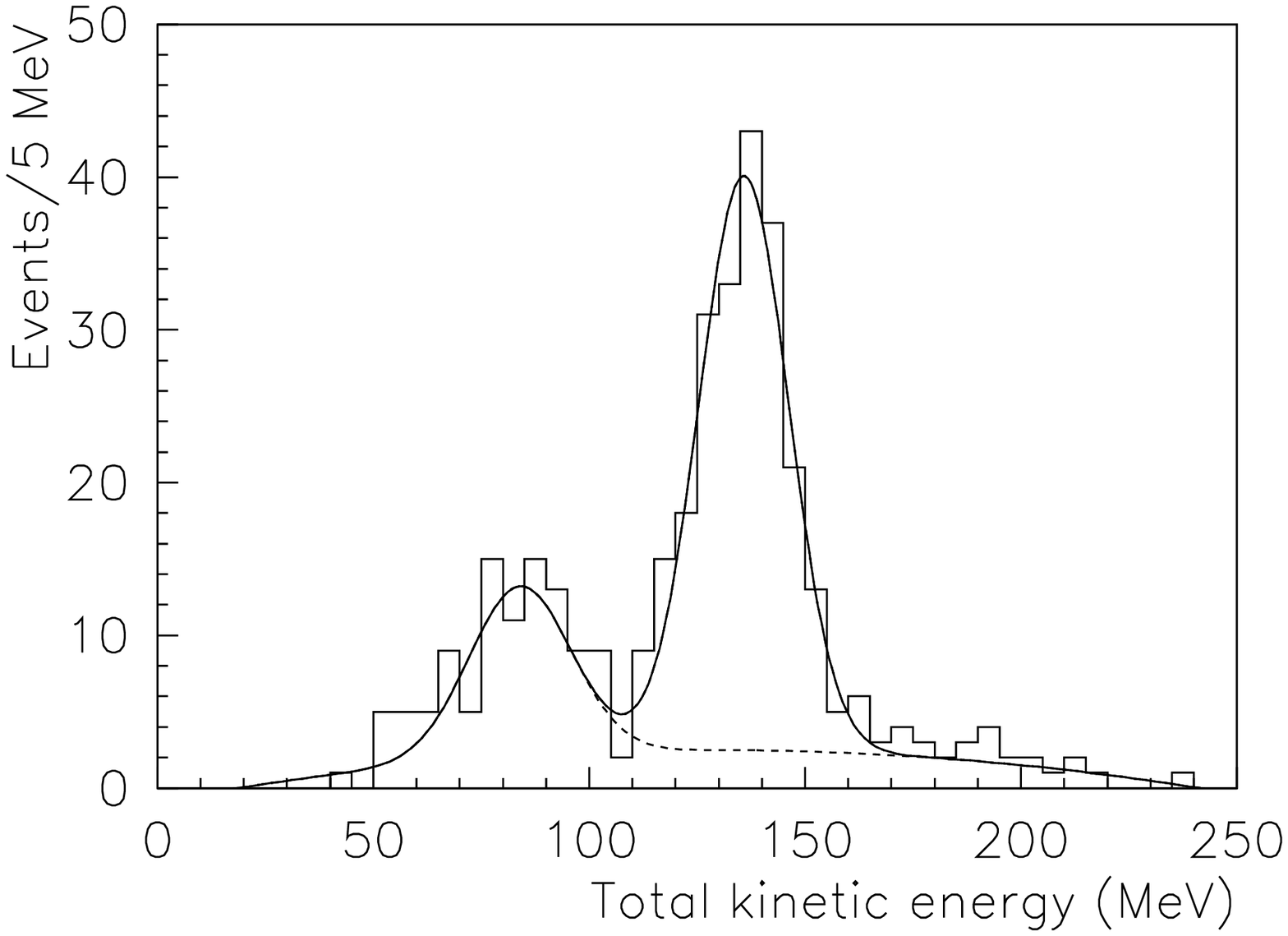}\hfil

\caption{Total decay-product kinetic energy of events 
surviving the two-track analysis.
The peak at 135 MeV corresponds to the expected kinetic energy in 
$K^{+}\!\rightarrow\!\pi^{+}\mu^{+}\mu^{-}$
decay.  The smooth curve is the fit to the spectrum with (solid) and
without (dashed) the Gaussian term for the signal.}
\label{mehran}
\end{figure}

The total kinetic energy distribution of the final two-track sample is shown
in Fig.~\ref{mehran}.  The large peak at about 135 MeV coincides with the
Monte Carlo distribution for $K^{+}\!\rightarrow\!\pi^{+}\mu^{+}\mu^{-}$.
The discrepancy between the peak position and the expected 143 MeV results
primarily from fibers in the target where energy deposits from the
kaon hide the energy deposited by decay particles.
Unrejected $K^{+}\!\rightarrow\!\pi^{+}\pi^{+}\pi^{-}$  events 
produce the smaller peak at 80 MeV.
The background extending into the 
$K^{+}\!\rightarrow\!\pi^{+}\mu^{+}\mu^{-}$ signal region is mostly
due to $K^{+}\!\rightarrow\!\pi^{+}\pi^{+}\pi^{-}$ decays with 
additional energy released in $\pi^-$ nuclear
capture. 
To estimate the sizes of the signal and background, a
function consisting of two Gaussians and a second-degree
polynomial is fitted to the spectrum (see Fig.~\ref{mehran}). 
The number of the signal events from the fit is $196.0\pm 16.7$.
The fitted number of background events in the interval 
$110<E_{kin}<160$ is 25.
Variations in the signal with the functional form assumed in the fit are
included in the systematic error discussed below.

To confirm identification of the main peak in Fig.~2 as 
$K^{+}\!\rightarrow\!\pi^{+}\mu^{+}\mu^{-}$, the TD's
are used to look for a $\pi^+\!\rightarrow\!\mu^+$ decay in the last counter
on each track.  
This allows an estimate of the number of $\pi^+\mu^+$ and
$\pi^+\pi^+$ pairs in the final sample in different regions of the
energy distribution.  
The fraction of $\pi^+\mu^+$ pairs in the signal region (110 to 160 MeV) 
is consistent 
with the ratio of fitted signal and background,
and the peak attributed to
$K^{+}\!\rightarrow\!\pi^{+}\pi^{+}\pi^{-}$ is mostly $\pi^+\pi^+$ pairs.
Monte Carlo studies show that the number of $\pi^+\mu^+$ pairs in the signal
region due to a decay-in-flight of one of the pions in a 
$K^{+}\!\rightarrow\!\pi^{+}\pi^{+}\pi^{-}$ decay
is negligible. The other $K^+$ decay involving a $\pi^+\mu^+$ pair is
$K^+\!\rightarrow\!\pi^+\pi^-\mu^+\nu$.
The maximum total kinetic
energy of the charged particles in this decay is 108.9 MeV
and the number of these
events in the signal region is estimated to be less than five.  

The number of stopped kaons is determined by normalizing a sample of 
$K^+\!\rightarrow\!\mu^+\nu$ events, taken simultaneously with the 
$K^{+}\!\rightarrow\!\pi^{+}\mu^{+}\mu^{-}$ sample, 
to the known branching ratio.
Several possible systematic errors in the final branching ratio cancel
via this procedure.
To estimate the acceptance for 
$K^{+}\!\rightarrow\!\pi^{+}\mu^{+}\mu^{-}$, Monte Carlo simulation is used for
kinematic factors and data are used for accidental losses, particle
identification, and timing 
requirements.
The simulation used the matrix element from
ChPT\cite{ecker} with the parameter 
$w_{+}=0.89$\cite{alliegro}.
Both analyses have acceptance in the region of dimuon invariant mass
from $m_{\mu\mu}=211$ MeV$/c^2$ (the endpoint) to $m_{\mu\mu}\approx 300$ 
MeV$/c^2$ (where the $\pi^+$ can barely reach the range stack).

The trigger efficiency varied from 5.3\% to 9.3\% over the course of
three years \cite{trigprob}.
The full acceptance for the trigger and three-track
analysis varied from 
$8.3 \times 10^{-4}$ to $9.6 \times 10^{-4}$ over the three years of data.
The branching ratio from the three track analysis is 
$(3.9 \pm 1.7^{\rm stat} \pm 0.4^{\rm syst})\times 10^{-8}$.
The estimated systematic error is dominated by uncertainty in the acceptance,
obtained by comparing a parallel measurement of
the $K^{+}\!\rightarrow\! \pi^{+}\pi^{-} e^{+} \nu$ 
branching ratio with the world average branching ratio.

The full acceptance of the two-track search, including trigger efficiency,
varied from 0.011 to 0.017, 
resulting in a branching ratio 
of $(5.1 \pm 0.4^{\rm stat} \pm 0.7^{\rm syst}) \times 10^{-8}$.
The systematic error includes the uncertainty in the background
shape, found by variations in the functional form used in the fit,
and a catalog of uncertainties affecting the acceptance,
the largest being the thickness of the trigger counters.

Taking into account the fact that the systematic errors in normalization 
and acceptance are correlated between the two analyses, we combine these two 
consistent results to give
$B(K^{+}\!\rightarrow\!\pi^{+}\mu^{+}\mu^{-})=
(5.0 \pm 0.4^{\rm stat} \pm 0.7^{\rm syst} \pm 0.6^{\rm th}) 
\times 10^{-8}$,
where the systematic uncertainty is the sum in quadrature of
0.5 (acceptance), 0.4 (background subtraction), and 0.2 (normalization)
\cite{overlap}.
The 12\% theoretical error comes from the effects of varying the shape of the
spectrum on the acceptance:  we vary $w_+$ in the ChPT
matrix element from -2 to +2 and note that the resulting range in
acceptance includes that for a pure vector coupling.

In the formalism of ChPT, the above result for
$B(K^+ \!\rightarrow\! \pi^+ \mu^+ \mu^-)$ implies $w_+ = 1.07\pm 0.07$
\cite{comment}.  
This can be compared with the values of $w_+$ extracted from 
$K^+ \!\rightarrow\! \pi^+ e^+ e^-$ described above.
We can also compare our result with the prediction of ChPT by using 
our branching ratio as well as the combined branching ratio 
and spectral shape fit of Ref. \cite{alliegro}.
We find $\Gamma(K^+ \!\rightarrow\! \pi^+
\mu^+ \mu^-)/\Gamma(K^+ \!\rightarrow\! \pi^+ e^+ e^-) = 0.167 \pm 0.036$, 
which is about $2.2 \sigma$ below the ChPT prediction of 0.236
at $w_+ =0.89$.
In the range $0.75 < w_+ < 1.13$ ($\pm 1\sigma$), this disagreement 
varies between 2.7 and $1.8 \sigma$.  
Note that the disagreement with the model of
Ref. \cite{except}, which predicts $\Gamma(K^+ \!\rightarrow\! \pi^+ \mu^+
\mu^-)/\Gamma(K^+ \!\rightarrow\! \pi^+ e^+ e^-) = 0.24$, is similar ($\sim 2
\sigma$).


\acknowledgments

We gratefully acknowledge the dedicated efforts of the technical staffs
supporting this experiment and of the Brookhaven AGS Department.
This research was supported in part by the U.S. Department of Energy
under contracts DE-AC02-76CH00016, W-7405-ENG-36, and 
grant DE-FG02-91ER40671
and by the Natural Sciences and Engineering Research Council and the
National Research Council of Canada.



\newpage

\begin{references}
\bibitem{idsamp} Two samples of $e^{\pm}$ are used: Dalitz decays of 
the $\pi^0$ from
$K^{+}\!\rightarrow\!\pi^{+}\pi^{0}$ decay
are tagged by the
205 MeV$/c$ $\pi^+$, and individual tracks making a complete orbit
in the central drift chamber are tagged as electrons using orbital time of 
flight.
The $\pi^+$ sample is tagged by its
$\pi^{+} \!\rightarrow\! \mu^{+} \nu$ decay,
observed in the range stack using the TD's.
\bibitem{trigprob} A trigger hardware problem caused the lower efficiency.
The full three-track acceptance shows less variation -- detector improvements
added a new particle identification tool in this time period that offset
the efficiency loss.
\bibitem{overlap} There is considerable overlap between the three-track events
found in the two analyses.  However, these represent a negligible contribution
to the full two-track analysis.
\bibitem{comment}We have added the statistical and systematic errors in 
                 quadrature for this calculation.  The changes in
                 acceptance from $w_+=0.89$ to $w_+=1.07$ are negligible.
                 We note that $w_+=-0.61\pm 0.07$ 
                 is also consistent with the value of 
       $B(K^{+}\!\rightarrow\!\pi^{+}\mu^{+}\mu^{-} )$
                 found here, but we 
                 assume that the spectral shape measured by \cite{alliegro} 
                 rules out the negative branch.
\end{references}
\end{document}